\def\bey{\begin{eqnarray}}
\def\eey{\end{eqnarray}}
\def\be{\begin{equation}}
\def\ee{\end{equation}}
\def\ba{\begin{array}}
\def\ea{\end{array}}

\def\af{\alpha}
\def\sg{\sigma}

\def\om{\omega}
\def\r{\rho}

\documentclass[onecolumn,showpacs,preprintnumbers,amsmath,amssymb]{revtex4}
\usepackage{graphicx}
\usepackage{fancyhdr}
\usepackage{dcolumn}
\usepackage{bm}
\usepackage{amssymb}
\usepackage{amsmath}
\usepackage{graphicx}
\usepackage[normalem]{ulem}
\usepackage[dvips]{color}

\begin{document}
\preprint{ }

\title{Neutron-skin thickness of finite nuclei in relativistic mean-field
models with chiral limits}
\author{Wei-Zhou Jiang$^{1,2}$, Bao-An Li$^{1}$, and Lie-Wen Chen$^{1,3}$}
\affiliation{  $^1$ Department of Physics, Texas A\&M
University-Commerce, Commerce, TX 75429, USA\\ $^2$ Institute of
Applied Physics, Chinese Academy of Sciences, Shanghai 201800,
China\\$^3$ Institute of Theoretical Physics, Shanghai Jiao Tong
University, Shanghai 200240, China}

\begin{abstract}
\baselineskip14pt We study several structure properties of finite
nuclei using relativistic mean-field Lagrangians constructed
according to the Brown-Rho scaling due to the chiral symmetry
restoration at high densities. The models are consistent with current
experimental constraints for the equations of state of symmetric
matter at both normal and supra-normal densities and of asymmetric
matter at sub-saturation densities. It is shown that these models can
successfully describe the binding energies and charge radii of finite
nuclei. Compared to calculations with usual relativistic mean-field
models, these models give a reduced thickness of neutron skin in
$^{208}$Pb between $0.17$ fm and $0.21$ fm. The reduction of the
predicted neutron skin thickness is found to be due to not only the
softening of the symmetry energy but also the scaling property of
$\rho $ meson required by the partial restoration of chiral symmetry.
\end{abstract}

\baselineskip 16pt
  \maketitle

\section{Introduction}

The nuclear symmetry energy of isospin asymmetric nuclear matter not
only plays a crucial role in a number of important issues in
astrophysics, see, e.g., Refs.~\cite{lat01,steiner05a}, but is also
important for understanding the structure of neutron- or proton-rich
nuclei and the reaction dynamics of heavy-ion collisions, see, e.g.,
Ref.~\cite{ba01,ho01,ji05}. However, the density dependence of the
symmetry energy is still poorly known especially at supra-normal
densities~\cite{li02,chen07}. In a previous work~\cite{ji07}, we
constructed several relativistic mean-field (RMF) Lagrangians using
in-medium hadron properties according to the Brown-Rho (BR) scaling
due to the chiral symmetry restoration at high
densities\cite{br91,br07}. The scalings and associated parameters
that describe the in-medium hadron properties are consistent with
those from microscopic calculations or those extracted from recent
experimental data. The symmetric part of the resulting equations of
state (EOS) around normal density is consistent with the data of
nuclear giant monopole resonances~\cite{You99} and at supra-normal
densities it is constrained by the collective flow data from high
energy heavy-ion reactions~\cite{da02}. Moreover, the density
dependence of the symmetry energy at sub-saturation densities is in
agreement with that extracted from the recent isospin diffusion data
from intermediate energy heavy-ion reactions~\cite{ts04,ch05,li05}.
An important feature of our models with the chiral limits is that the
resulting symmetric nuclear matter EOS is soft at intermediate
densities but stiff at high densities naturally, producing a maximum
neutron star mass around 2.0$M_\odot$ consistent with the recent
astrophysical observations.

In the present work, we extend our well constrained RMF models to
study ground-state properties of finite nuclei. In particular, we
examine the neutron skin thickness of $^{208}$Pb. The size of the
neutron skin of finite nuclei is determined by the competition
between the neutron pressure and the surface tension. Indeed, it has
been shown in many studies that the neutron skin thickness of
$^{208}$Pb is rather sensitive to the density dependence of the
symmetry energy~\cite{brown00,typel01,pi02,vr03,che05,furn02}. The
latter also influences the extraction of the incompressibility of
nuclear matter from the experimental data of giant resonances in
finite nuclei~\cite{vr03,pik04,colo04}. The size of the neutron-skin
in $^{208}$Pb predicted by these models thus also appears to be
limited by both the symmetry energy and the nuclear matter
incompressibility of the models used. For instance, non-relativistic
models that give an incompressibility of about $\kappa=220\pm15$MeV
and a symmetry energy at normal density less than 32 MeV produce a
neutron skin thickness for $^{208}$Pb smaller than 0.2 fm, while the
RMF models that have larger incompressibilities of 250-270 MeV and
the symmetry energy at normal density larger than 34 MeV predict a
neutron skin thickness for $^{208}$Pb as large as about
0.3fm~\cite{furn02}. However, it is understood that it is the slope
of the underlying symmetry energy that matters the most\cite{furn02}.
It was pointed out by Vretenar et.al.~\cite{vr03} that the
incompressibility in RMF models was necessarily larger than 250 MeV
for reproducing the nuclear structural properties. In a newly
developed RMF model  with the softened density dependence of the
symmetry energy, Todd-Rutel and Piekarewicz  obtained the
incompressibility 230 MeV  and predicted a neutron skin thickness of
0.21fm for $^{208}$Pb~\cite{fsu}. On one hand, our models have the
characteristic of  a small symmetry energy of 31.6 MeV  and a small
incompressibility of 230MeV at saturation density
$\rho_0=0.16$fm$^{-3}$. One would thus expect a thin neutron skin in
$^{208}$Pb. On the other hand, the cancellation of the medium effect
from the vector meson mass and its coupling constant in our models
results in an EOS in  resemblance to those of the linear Walecka
model and the best-fit model NL3 which have a much larger neutron
skin thickness in $^{208}$Pb. Given the successful description of the
high-density behavior of nuclear matter, it is necessary to extend
our models to the low density region by studying the ground-state
properties of finite nuclei, especially the neutron skin thickness of
heavy nuclei. Moreover, it is important to have a better
understanding about the various contributing factors to the neutron
skin thickness within the RMF models.

The paper is organized as follows. In Section \ref{RMF}, we
briefly introduce the formalism of RMF with Brown-Rho scaling due
to the chiral symmetry restoration at high densities. Results on
the ground-state properties of finite nuclei, especially the
neutron-skin thickness of finite nuclei are presented in Section
\ref{results}. A summary is finally given in Section
\ref{summary}.

\section{A brief summary of the formalism}
\label{RMF}

A lot of efforts had been devoted to describing well the binding
energies and charge radii of finite nuclei simultaneously\cite%
{chi74,bo77,ho81,re86} within the RMF models. Inspired by the
nonlinearity of $\sigma$ meson in chiral $\sigma$ model or induced
by the model renormalization, nonlinear $\sigma$ meson terms were
introduced to describe the medium effects that are important in
the Brueckner theory. As a result, the binding energies and charge
radii of finite nuclei were well reproduced, along with an
improved description for the nuclear surface. Later on, the
$\omega$ meson nonlinear term was introduced in compliance with
the relativistic Brueckner results~\cite{tm1}. Some prominent
representatives of the best-fit RMF models are NL3~\cite{nl3},
TM1~\cite{tm1}, FSUGold~\cite{fsu}, and so on. Based on the
microscopic relativistic Brueckner results, the density-dependent
Hartree or Hartree-Fock approaches were developed to describe the
finite nuclei (see ~\cite{ma02,hof01} and references therein).
These approaches are usually parameter free but their predictions
are not as accurate as the best-fit models. Our models used here
have no nonlinear meson self-interactions, while the medium
effects are given by the BR scaling.

In the present work, the model Lagrangian with the
density-dependent couplings and meson masses is written as
\begin{eqnarray}
{\cal L}&=& {\overline\psi}[i\gamma_{\mu}\partial^{\mu}-M
+g^*_{\sigma}\sigma-g^*_{\omega }
\gamma_{\mu}\omega^{\mu}-g^*_{\rho}\gamma_\mu \tau_3 b_0^\mu-e\frac{1}{2}%
(1+\tau_3)\gamma_\mu A^\mu]\psi +\frac{1}{2}(\partial_{\mu}\sigma\partial^{%
\mu}\sigma-m_{\sigma}^{*2}\sigma^{2})  \nonumber \\
&& - \frac{1}{4}F_{\mu\nu}F^{\mu\nu}+ \frac{1}{2}m_{\omega}^{*2}\omega_{\mu}%
\omega^{\mu} - \frac{1}{4}B_{\mu\nu} B^{\mu\nu}+ \frac{1}{2}m_{\rho}^{*2}
b_{0\mu} b_0^{\mu}- \frac{1}{4}A_{\mu\nu} A^{\mu\nu}  \label{eq:lag1}
\end{eqnarray}
where $\psi,\sigma,\omega$, and $b_0$ are the fields of the nucleon, scalar,
vector, and isovector-vector mesons, with their masses $M, m^*_\sg,m^*_\om$,
and $m^*_\r$, respectively. The meson coupling constants and masses with
asterisks denote the density dependence, given by the BR scaling~\cite%
{ji07,br05,song01}. The Dirac equation of nucleons in the mean-field
approximation is given by
\begin{equation}
[i\gamma_{\mu}\partial^{\mu}-M +g^*_{\sigma}\sigma-g^*_{\omega }
\gamma_{0}\omega^{0}-g^*_{\rho}\gamma_0 \tau_3 b_0-e\frac{1}{2}%
(1+\tau_3)\gamma_0 A^0+\gamma_0\Sigma^{R,0}]\psi=0.  \label{eqn1}
\end{equation}
The rearrangement term $\Sigma^{R,0}$ is essential for the thermodynamic
consistency to derive the pressure and plays role in modifying the single
particle energy and total binding energy of finite nuclei. In the mean field
approximation, the $\Sigma^R_0$ is given by
\begin{equation}  \label{rre}
\Sigma^R_0=<\frac{\partial{\cal L}}{\partial\rho}>=m_\om^*\omega_0^2\frac{%
\partial m_\om^*}{\partial \rho} + m_\r^* b_0^2\frac{\partial m_\r^*}{%
\partial \rho}- m_\sg^*\sigma_0^2\frac{\partial m_\sg^*}{\partial \rho}
-\rho\omega_0\frac{\partial g_\om^*}{\partial \rho} -\rho_3 b_0\frac{%
\partial g_\r^*}{\partial \rho} +\rho_s\sigma\frac{\partial g_\sg^*}{%
\partial \rho}.
\end{equation}
The density dependence of parameters is described by the scaling functions
that are the ratios of the in-medium parameters to those in the free space.
We take the scaling functions for the coupling constants of scalar and
vector mesons as~\cite{ji07}
\begin{equation}  \label{sc3}
\Phi_\sg(\rho)=\frac{1}{1+x\rho/\rho_0},\hbox{ } \Phi_\r(\rho)=\frac{%
1-y\rho/\rho_0}{1+y_\r\rho/\rho_0}, \hbox{ } \Phi_\om(\rho)=\frac{%
1-y\rho/\rho_0}{1+y_\om\rho/\rho_0}.
\end{equation}
For hadron masses, the scaling function reads
\begin{equation}
\Phi(\rho)=1-y\rho/\rho_0.  \label{sc2}
\end{equation}
The energy per nucleon is given by~\cite{fu95}
\begin{eqnarray}  \label{eqe1}
E=\sum_\af(2j_\af+1)\epsilon_\af -\frac{1}{2}\int d^3\!r
(-g_\sg^*\sigma\rho_s+ g_\om^*\omega\rho +g_\r^*b_0\rho_3
+eA_0\rho_p-2\Sigma^R_0\rho)+E_{c.m.}
\end{eqnarray}
where $\epsilon$ is the eigen energy of nucleon, $\alpha$ denotes the
occupied state, and $j$ is the quantum number of total angular
momentum.
The energy from the center of mass correction is $E_{c.m.}=-3/4%
\times41A^{-1/3}$~\cite{re86}. The equation of motion of mesons can be
written out according to the Euler-Lagrange equation. The equations of
nucleons and mesons are coupled nonlinearly and can be solved by iterations.
The detailed procedure can be found in numerous literatures. Here, we just
emphasize that the meson mass in the Green function is that in the free
space. As an example, the $\sigma$ meson field is integrated out as follows
\begin{equation}  \label{eqsg}
\sigma(r)=\int d\!r_1 r_1^2 G(r,r_1,m_\sg)[-g_\sg^*(r_1)\rho_s(r_1)-
(m_{\sigma}^2-{m_\sg^*}^2(r_1))\sigma(r_1)]
\end{equation}
where $m_\sg$ is the free $\sigma$ meson mass,
$g^*_\sg(r_1)=g^*_\sg(\rho(r_1))$, and
$m^*_\sg(r_1)=m^*_\sg(\rho(r_1))$. The expression of $G$ can be found
in Ref.~\cite{ji06}.

\section{Results and discussions}
\label{results}

\begin{table}[tbp]
\caption{Parameter sets used in the calculations. The vacuum hadron masses
are $M=938$MeV, $m_\om=783$MeV and $m_\r=770$MeV. The coupling constants
given here are those at zero density. The incompressibility and the symmetry
energy are respectively 230MeV and 31.6MeV at saturation density $\protect%
\rho_0=0.16$fm$^{-3}$ for all models. The binding energy per nucleon (B/A), $%
m_\sg$, and the incompressibility $\protect\kappa$ are in unit of MeV. }
\label{t:t1}
\begin{center}
\begin{tabular}{cccccccccccc}
\hline\hline Model & $m_\sg$ & $g_\sg$ & $g_\om$ & $g_\r$ & $y$ & $x$
& $B/A$ & $M^*/M$ & $y_\r$ & $y_\om$\\ \hline
 SL1 & 600 & 10.3665 &10.4634 & 3.7875 & 0.126 & 0.234 & -16.0 & 0.679 & - &
-  \\
SLC & 590 & 10.1408 & 10.3261 & 3.8021 & 0.126 & 0.239 & -16.3 & 0.685 & - &
-  \\
SLCd & 590 & 10.1408 & 10.3261 & 5.7758 & 0.126 & 0.239 & -16.3 &
0.685 & 0.5191 &  -  \\ \hline\hline
\end{tabular}%
\end{center}
\end{table}

The details in determining model parameters can be found in our
previous work~\cite{ji07}. Here we will firstly do calculations
based on the parameter set SL1. In Table~\ref{t:t1}, we tabulate
the parameter set SL1 which is a little different from that in
Ref.~\cite{ji07} because of a larger $m_\sg$ here. The nuclear
matter property does not change at all by the variation of $m_\sg$
because the ratio $g_\sg/m_\sg$ keeps unchanged, while the
description for finite nuclei is modified. The latter can be seen
in Eq. (\ref{eqsg}) and is correct for all cases irrespective of
whether the coupling constant and meson mass are density dependent
or not. For the linear Walecka model, the binding energies and
charge radii can not be described well simultaneously by adjusting
the $m_\sg$~\cite{ho81}. The second term in the integrand of Eq.
(\ref{eqsg}) is of prime importance in improving the description
of the binding energies and charge radii of finite nuclei.
Besides, another improvement comes from the inclusion of the
rearrangement term appearing in Eqs. (\ref{eqn1}) and
(\ref{eqe1}). Using $m_\sg=500$MeV in the original version of SL1,
the total binding energy for $^{40}$Ca is smaller than the
experimental value by 70 MeV, and for $^{208}$Pb the deviation is
about 230MeV. These results are improved appreciably by shifting
the $m_\sg$ up to 600MeV, as shown in Table~\ref{t:t2}. The
relative error of charge radii is less than 1.5\% and that of
binding energy is less than 3\%.

\begin{figure}
\begin{center}
\vspace*{-10mm}
\includegraphics[width=10cm,height=10cm]{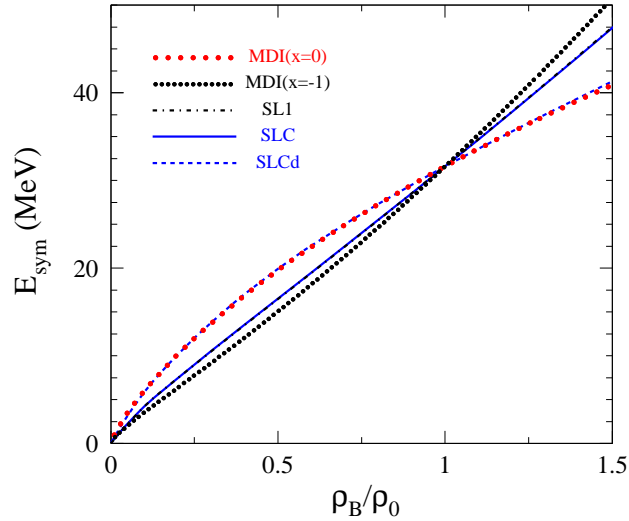}
 \end{center}
\caption{(Color online) The symmetry energy as a function of density
for different models. The SL1 and SLC results overlap completely.}
\label{f:f1}
\end{figure}

The accuracy of predictions can be much improved by setting $B/A=-16.3$MeV
and $m_\sg=590$MeV. Such a slight shift changes very little properties of
infinite nuclear matter since the incompressibility is kept the same as that
of the SL1 parameter set. The new parameter set called SLC is also listed in
Table~\ref{t:t1}. Except for $^{16}$O, excellent agreement with the
experimental data is achieved using the model SLC, as shown in Table~\ref%
{t:t2}, though the binding energy of medium-mass nuclei is slightly
underestimated by about 1.5\%. To see the effects of the density
dependence of the symmetry energy on the properties of finite nuclei,
we derive the parameter set SLCd. The parameter set SLCd is different
from the SLC only for the $\rho$ meson coupling constant and its
density dependence which is modified by an additional coefficient
$y_\r$, as given in Table~\ref{t:t1}. The results obtained by using
the SLCd parameter set are given in Table~\ref{t:t2}, where for
comparison the FSUGold results are also listed. Fig.~\ref{f:f1} shows
the density dependence of the symmetry energy with the SLC, SLCd and
SL1 parameter sets. For comparison, we also include the MDI(x=0) and
MDI(x=-1) results which represent the experimental constraints on the
density dependence of the symmetry energy at sub-saturation
densities, derived by studying the isospin diffusion data within a
transport model~\cite{ch05,li05}. As shown in Fig.~\ref{f:f1}, no
obvious difference between the SLC and SL1 is visible for the density
dependence of the symmetry energy. Interestingly, all the symmetry
energies from the parameter sets SLC, SLCd and SL1 are consistent
with that constrained by the isospin diffusion data.

\begin{table}[tbh]
\caption{The binding energy per nucleon (B/A), charge radii $r_c$ in fm, and
neutron skins $r_n-r_p$ obtained from different models. The available
experimental data are from \protect\cite{au03,an04}. }
\label{t:t2}
\begin{center}
\begin{tabular}{ccccccc}
\hline\hline
Nucleus &  & Expt. & SL1 & SLC & SLCd & FSUGold \\ \hline
$^{16}$O & B/A & 7.98 & 8.03 & 8.07 & 8.07 & 7.96 \\
& $r_c$ & 2.70 & 2.72 & 2.74 & 2.74 & 2.69 \\
& $r_n-r_p$ & - & -0.03 & -0.03 & -0.03 & -0.03 \\
$^{40}$Ca & B/A & 8.55 & 8.43 & 8.54 & 8.54 & 8.54 \\
& $r_c$ & 3.48 & 3.43 & 3.45 & 3.44 & 3.42 \\
& $r_n-r_p$ & - & -0.05 & -0.05 & -0.05 & -0.05 \\
$^{48}$Ca & B/A & 8.67 & 8.41 & 8.52 & 8.46 & 8.58 \\
& $r_c$ & 3.47 & 3.45 & 3.46 & 3.47 & 3.44 \\
& $r_n-r_p$ & - & 0.20 & 0.20 & 0.18 & 0.20 \\
$^{90}$Zr & B/A & 8.71 & 8.44 & 8.59 & 8.56 & 8.68 \\
& $r_c$ & 4.26 & 4.22 & 4.23 & 4.23 & 4.25 \\
& $r_n-r_p$ & - & 0.10 & 0.10 & 0.08 & 0.09 \\
$^{114}$Sn & B/A & 8.52 & 8.24 & 8.40 & 8.38 & 8.49 \\
& $r_c$ & 4.61 & 4.57 & 4.58 & 4.59 & 4.60 \\
& $r_n-r_p$ & - & 0.09 & 0.09 & 0.07 & 0.09 \\
$^{132}$Sn & B/A & 8.36 & 8.18 & 8.35 & 8.23 & 8.34 \\
& $r_c$ & - & 4.68 & 4.69 & 4.70 & 4.71 \\
& $r_n-r_p$ & - & 0.27 & 0.27 & 0.23 & 0.27 \\
$^{208}$Pb & B/A & 7.87 & 7.68 & 7.87 & 7.79 & 7.89 \\
& $r_c$ & 5.50 & 5.46 & 5.47 & 5.48 & 5.52 \\
& $r_n-r_p$ & - & 0.21 & 0.21 & 0.17 & 0.21 \\ \hline\hline
\end{tabular}%
\end{center}
\end{table}

It is now well known that the neutron skin thickness ($r_n-r_p$) of
heavy nuclei depends sensitively on the slope of the symmetry energy
at normal density. A stiffer density dependence of the symmetry
energy results in a thicker neutron skin. The difference between the
symmetry energies with the SLC and SLCd, shown in Fig.~\ref{f:f1}, is
responsible for the variation of the neutron skin in neutron-rich
nuclei as seen in Table~\ref{t:t2}. Noting that the model FSUGold has
almost the same symmetry energy in the whole density region as that
of the MDI(x=0) and has larger symmetry energies at low densities
than the SL1 and SLC, the neutron-skin thickness in $^{208}$Pb with
the SL1 and SLC would thus be expected to be much larger than that
given by the FSUGold. However, from Table~\ref{t:t2}, we can see
surprisingly that the SL1, SLC and FSUGold predict almost the same
neutron skin for magic nuclei listed in Table~\ref{t:t2}. Therefore,
given the same density dependence of the symmetry energy as with the
best-fit model FSUGold, the present models with the BR scaling
predict smaller values for the neutron-skin thickness for finite
nuclei. Compared to other best-fit RMF models, such as the NL1 and
TM1 that predict the thickness of neutron skin in $^{208}$Pb to be
0.28fm and 0.26fm, respectively, this reduction of the predicted
neutron-skin thickness is more appreciable. To understand the
underlying physics for the above observation, we may write down the
integration form for the isovector potential that is a product of the
$\rho$ meson field and the coupling constant $g^*_\r$
\begin{equation}  \label{eqrho}
V_\r(r)=g^*_\r(r)\int d\!r_1 r_1^2 G(r,r_1,m_\r)
[-g_\r^*(r_1)\rho_3(r_1)-(m_{\rho}^2-{m_\r^*}^2(r_1))b_0(r_1)].
\end{equation}
For the SL1 and SLC, the ratio $C_\r=g_\r^*/m_\r^*$ is a constant.
Properties of asymmetric nuclear matter in the RMF approximation
depend on the constant ratio $C_\r$ rather than the respective values
of the coupling constant and the mass. This is the same as in the
linear Walecka model and the best-fit model NL3. However, the linear
Walecka model and the NL3 model predict much thicker neutron skin of
0.27fm and 0.28fm in $^{208}$Pb, respectively~\cite{ho81,nl3}. As the
symmetry energy changes from 35MeV to 31.6MeV in the linear Walecka
model, the neutron thickness is still as large as 0.25fm. It is
similar for the NL3 model, but a reduction in the symmetry energy to
31.6MeV results in large deviations of calculated nucleus masses from
the experimental values~\cite{vr03}, which is strongly unfavored by
the best-fit model. The linear Walecka model, the NL3 and the FSUGold
all have a much smaller effective nucleon mass $M^*$ at the
saturation density than ours. It should be noted that it is not the
larger $M^*$ that results in the small neutron skin since a small
neutron skin of 0.22fm in $^{208}$Pb is still obtained with our model
that is adjusted to have a small $ M^*=0.6M$ with the same symmetry
energies as given by the SL1 and SLC parameter sets. It is
interesting that the scaling property of the $\rho$ meson that is
hidden behind calculations for nuclear matter is now displayed
clearly in finite nuclei. The second term in the integrand of Eq.
(\ref{eqrho}) is induced by the $\rho$ mass scaling. Together with
the density dependent coupling constant, it is responsible for the
reduction of the neutron skin thickness in neutron-rich nuclei. The
drop of the neutron skin thickness with the SLCd, compared to the SL1
and SLC, can just be attributed to the softening of the symmetry
energy. We note that the neutron skin is totally insensitive to the
value of $m_\sg$, though a better fit of both binding energies and
charge radii relies on the choice of $m_\sg$. The thickness of
neutron skin in $^{208}$Pb varies from 0.21fm to 0.17fm by changing
from the SLC to the SLCd. This range is consistent with current
measurements using the X-ray cascade from antiprotonic atoms:
$0.16\pm0.06$fm~\cite{kl07} and also agrees well
with that obtained from the analysis of the isospin diffusion data~\cite%
{che05}: $0.22\pm0.04$. Interestingly, the predicted range of neutron skin
in $^{208}$Pb between 0.21fm and 0.17fm is very close to that predicted by
using some Skyrme interactions~\cite{furn02}.

It is worth noting that a much smaller neutron-skin thickness in
$^{208}$Pb can be obtained within the RMF models provided that
smaller values of   symmetry energy are given. For instance, the ES25
model~\cite{steiner05a} constructed with the incompressibility of
211.7 MeV and the symmetry energy of 25 MeV  at saturation density
predicts a neutron skin as thin as 0.138 fm for $^{208}$Pb. With the
same incompressibility and symmetry energies as the ES25 model, our
model predicts an even smaller neutron skin of 0.126 fm for
$^{208}$Pb. Of course, the magnitude of the reduction due to the
in-medium properties of the $\rho$ meson is accordingly smaller.

The deficiency of our models mainly lies in the unsatisfactory
description for spin-orbit splittings. For instance, the $1p$
spin-orbit splitting in $^{16}$O is 3.7 and 3.5MeV with the SL1 and
SLC, respectively, which is smaller than the experimental value by
about 2.5 MeV. This deficiency stems mainly from a much larger Dirac
mass $M^*$ in our present models, i.e., about 0.68M. Usually, a Dirac
mass of about 0.5-0.6M can reasonably
describe the spin-orbit splitting in RMF models (see, e.g., Ref.~\cite%
{typel05}). As a result, the spin-orbit potential of the central part
is much suppressed in our models, compared to that given by the
best-fit models. The deviation of spin-orbit splittings from the
available data is comparable to that given by the density dependent
approaches based on the relativistic Brueckner results~\cite{hof01}.
The right spin-orbit splitting in our calculations can be obtained by
reducing the Dirac mass moderately. On the other hand, the larger
Dirac mass leads to larger Landau mass (about 0.74 in our models),
which gives a more reasonable description for the level
density~\cite{typel05}.

\section{Summary}
\label{summary}

In summary, we have studied ground-state properties of finite nuclei
within the relativistic mean-field models that are constructed
according to the Brown-Rho scaling due to the chiral symmetry
restoration at high densities. Not only can the constructed models
SLC and SLCd be consistent with current experimental results for the
equation of state of symmetric matter at normal and supra-normal
densities and of asymmetric matter at sub-saturation densities, but
also present a fairly satisfactory description for the ground state
properties of finite nuclei. The binding energies and charge radii of
a variety of magic nuclei are nicely reproduced. Our results indicate
that the neutron-skin thickness of finite nuclei depends not only on
the density dependence of the symmetry energy but also on the scaling
property of the $\rho$ meson. Compared to  calculations of usual RMF
models a reduction of neutron skin thickness of neutron-rich nuclei
is observed in our models. We find that the scaling property of the
$\rho$ meson that is hidden in nuclear matter calculations is shown
up clearly by the reduction of the neutron skin in $^{208}$Pb. The
neutron skin thickness of $^{208}$Pb is predicted to be between 0.17
fm to 0.21 fm consistent with the current measurements.

\section*{Acknowledgement}

The work was supported in part by the US National Science Foundation
under Grant No. PHY-0652548, the Research Corporation under Award
No. 7123, the National Natural Science Foundation of China under
Grant Nos. 10405031, 10575071 and 10675082, MOE of China under
project NCET-05-0392, Shanghai Rising-Star Program under Grant No.
06QA14024, the SRF for ROCS, SEM of China,  the Knowledge Innovation
Project of the Chinese Academy of Sciences under Grant No.
KJXC3-SYW-N2, and the China Major State Basic Research Development
Program under Contract No. 2007CB815004.

\end{document}